\documentclass[twocolumn]{autart}    % Enable this line and disable the
\usepackage{mathrsfs}
\usepackage{float}
\usepackage{amsfonts}
\usepackage{graphicx}          % Include this line if your
\usepackage{color}
\usepackage{amsmath}
\usepackage{caption}
\usepackage[titletoc,title]{appendix}
\usepackage[round]{natbib}
\usepackage{diagbox}
\usepackage{makecell}
\usepackage{booktabs}

\newtheorem{remark}{Remark}

\newtheorem{lemma}{Lemma}

\newtheorem{proof of theorem 2}{Proof of Theorem 2}
\newtheorem{proof of theorem 1}{Proof of Theorem 1}

\begin{document}
\hyphenpenalty=5000
\tolerance=1000
\begin{frontmatter}

\title{Event-Triggered Extended State Observer Based Distributed Control of Nonlinear Vehicle Platoons} % Title, preferably not more

\thanks[footnoteinfo]{
Corresponding author: Tao Li. Tel. +86-21-54342646-318.
Fax +86-21-54342609.}
\author[Paestum]{Anquan Liu}\ead{anquanliu186@163.com},    % Add the
\author[Rome]{Tao Li}
%\thanks{* Corresponding author}
\ead{tli@math.ecnu.edu.cn}\textsuperscript{,*},               % e-mail address
\author[Dome]{Yu Gu}\ead{jessicagyrrr@126.com},  % (ead) as shown

\address[Paestum]{School of Mechatronic Engineering and Automation, Shanghai University, Shanghai, 200072, China.}  % Please supply
\address[Rome]{Shanghai Key Laboratory of Pure Mathematics and Mathematical Practice,
	School of Mathematical Sciences, East China Normal University, Shanghai
	200241, China.}             % full addresses
\address[Dome]{Shanghai Municipal Educational Examinations Authority, Shanghai, 200433, China.}

\begin{keyword}                           % Five to ten keywords,
Platoon control; Event-triggered extended state observer;  Modified dynamic surface control; String stability             % chosen from the IFAC
\end{keyword}                             % keyword list or with the% help of the Automatica
\begin{abstract}                          % Abstract of not more than 200 words.
We study the platoon control of vehicles with third-order nonlinear dynamics under the constant spacing policy. We consider a vehicle model with parameter uncertainties and external disturbances and propose a distributed control law based on an event-triggered extended state observer (ESO). First, an event-triggered ESO is designed to estimate the unmodeled dynamics in the vehicle model. Then based on the estimate of the unmodeled dynamics, a distributed control law is designed by using a modified dynamic surface control method. The control law of each follower vehicle only uses the information obtained by on-board sensors, including its own velocity, acceleration, the velocity of the preceding vehicle and the inter-vehicle distance. Finally, we give the range of the control parameters to ensure the stability of the vehicle platoon system. It is shown that
the control parameters can be properly designed to make the observation errors of the ESOs bounded and ensure the string stability and closed-loop stability. We prove that the Zeno behavior is avoided under the designed event-triggered mechanism. The joint simulations of CarSim and MATLAB are given to demonstrate the effectiveness of the proposed control law.
\end{abstract}

\end{frontmatter}

\section{Introduction}

Vehicle platoon control has attracted widespread attention due to its many advantages, such as enhanced road safety, high road utilization, low energy consumption and so on. The early research on vehicle platoon control can be traced back to the California Partners for Advanced Transit and Highways (PATH) program in 1986. In the past decades, the platoon control research community has achieved fruitful results \citep{guanetti2018control}.

Many factors need to be considered in the vehicle platoon control, such as spacing policy, inter-vehicle information flow topology, vehicle dynamics and so on \citep{li2015overview}. The vehicle dynamics model can be divided into two types: linear and nonlinear, and many researchers have considered the former one
\citep{di2014distributed,hao2013stability,jovanovic2005ill,eyre1998simplified,sheikholeslam1993longitudinal,naus2010string,xiao2011practical,oncu2014cooperative,zheng2015stability,guo2012autonomous}.
Although it is convenient to carry on theoretical analysis by using linear vehicle models, the vehicle in reality is a complex nonlinear dynamic system. The dynamic characteristics of vehicles cannot be fully described by simple linear models.
The control law designed based on the linear vehicle model may not achieve good control effect for the vehicle with nonlinear dynamics.
 More and more researchers have begun to consider the nonlinear vehicle platoon control.
 Nonlinear vehicle models with known parameters are considered by \cite{wu2016distributed} and \cite{zheng2016distributed}. Nonlinear vehicle models with parameter uncertainties are considered by \cite{yue2012guaranteed}, \cite{zhu2018distributed}, \cite{chehardoli2018adaptive} and \cite{zhu2019barrier}.  Nonlinear vehicle models with parameter uncertainties and unknown external disturbances are considered by \cite{guo2017cnn}, \cite{kwon2014adaptive}, \cite{guo2017adaptive} and \cite{li2020prescribed}.

The adaptive control scheme is often used to deal with parameter uncertainties and external disturbances in the vehicle model. Some researchers designed adaptive control laws to estimate the unknown parameters and the bounds of the external disturbances, however, the vehicle models are required to be parametric linearized for adaptive control laws. Neural networks are usually utilized to approximate the unmodeled dynamics in the vehicle model, but the structure of neural networks usually needs to be determined empirically and a large
number of parameters need to be accordingly designed. In addition, the control laws in the above literature rely on the information obtained by the wireless communication network, such as the acceleration of the preceding vehicle, the velocity and the acceleration of the leader vehicle. Although the rapid development of the wireless communication technology diversifies information flow topologies, the issues brought by the wireless communication can not be ignored, such as time delays, packet losses, network attacks  and so on \citep{willke2009survey}.  The distributed control law in \cite{liu2021cooperative} which only relies on the information obtained by on-board sensors can ensure the stability of the vehicle platoon system for any given positive time headway, however, \cite{liu2021cooperative} considered a linear vehicle model.

\vspace{-2mm}
In this paper, we study the platoon control of vehicles with third-order nonlinear dynamics under the constant spacing policy. We consider vehicle models with parameter uncertainties and external disturbances. Firstly, an event-triggered ESO is designed to estimate the unmodeled dynamics in the vehicle model. Then based on the estimate of the unmodeled dynamics, we present a distributed control law through a modified dynamic surface control method. In the framework of the ESO, the  accurate vehicle model is not required and the linear parameterization of the vehicle model is avoided. At the same time, ESOs have a simple structure and few parameters. On-board systems are usually powered by batteries and data transmission is energy-consuming \citep{miskowicz2018event,ge2021dynamic}. In order to reduce the energy consumption caused by the data transmission,
an event-triggered mechanism is embedded in the ESO, which effectively reduces the data transmission from the controller to the ESO.
In order to ensure the string stability, a virtual velocity gain and a virtual acceleration gain are introduced into the dynamic surface control method, which can be adjusted to guarantee the string stability. The modified dynamic surface control method can avoid the derivation of the virtual control input, thereby the acceleration of the preceding vehicle which is usually obtained by the wireless communication is not needed by the control law. The control law of each follower vehicle is only based on its own velocity, acceleration, the velocity of the preceding vehicle and the inter-vehicle distance, which are all obtained by on-board sensors.

\vspace{-1mm}
We analyze the stability of the vehicle platoon system. Firstly, the boundedness of observation errors of the event-triggered ESOs is analyzed.
We prove that the observation errors of the ESOs are bounded under the designed event-trigger mechanisms and give the upper bounds of the observation errors. Then on the basis of the bounded observation errors, we give the explicit range of the control parameters to ensure the stability of the vehicle platoon system. It is shown that the control parameters can be properly designed to ensure the string stability and closed-loop stability of the vehicle platoon system. We also prove that there exists a positive lower bound of the time interval of the event triggering, therefore, the Zeno behavior is avoided under the designed event-triggered mechanism.

Some numerical simulations are given to demonstrate the effectiveness of the proposed control laws. Numerical simulations consist of two parts. In the first part, the nonlinear vehicle models given in this paper are considered in the simulations. In the second part, the vehicle models in CarSim are considered and the joint simulations of Simulink and CarSim are given. It is shown that although the control laws are designed for the simplified nonlinear vehicle models, they can still achieve good control effect for the vehicle models in CarSim which are more consistent with the dynamic characteristics of real vehicles.

The rest of this paper is organized as follows. The vehicle models and the control objectives are given in Section 2. In Section 3, we present a distributed control law based on an  event-triggered ESO and a modified dynamic surface control method. In Section 4, we analyze the stability of the vehicle platoon system and give the range of the control parameters to ensure the string stability and the closed-loop stability. Numerical simulations are performed in Section 5. Some conclusions and future research topics are given in Section 6.
\vspace{-2mm}
\section{Problem formulation}
\vspace{-2mm}
Suppose that there are $N+1$ vehicles in the platoon, including one virtual leader vehicle and $N$ follower vehicles. Consider the following virtual leader vehicle model
\begin{equation}\label{Model0}
\left\{
\begin{aligned}
\dot p_0(t) &= v_0(t), \vspace{1ex} \\
\dot v_0(t) &= a_0(t), \vspace{1ex} \\
\dot a_0(t) &= -a_0(t)/\tau_0+u_0(t)/\tau_0,
\end{aligned}
\right.
\end{equation}
and follower vehicle models
\begin{equation}\label{Model}
\left\{
\begin{aligned}
\dot p_i(t) = &v_i(t), \vspace{1ex} \\
\dot v_i(t) = &a_i(t),  \hspace{30mm} i = 1,2,\ldots,N,\\
\dot a_i(t) = &-a_i(t)/\tau_i-c_i v_i^2(t)/(m_i\tau_i)-g\mu_i/\tau_i\\
              &-2c_iv_i(t)a_i(t)/m_i+u_i(t)/(m_i\tau_i)+\sigma_i(t),
\end{aligned}
\right.
\end{equation}
where $p_i(t)$, $v_i(t)$, $a_i(t)$ are the position, velocity and acceleration of the vehicle $i$, respectively, $i=0,1,...,N$. The constant $\tau_0$ is the inertial delay of the leader vehicle, and $u_0(t)$ is the externally given control input of the leader vehicle. The constants $m_i$, $c_i$, $\mu_i$ and $\tau_i$ are the mass, total air resistance coefficient, rolling resistance coefficient and inertial delay of the $i$th follower vehicle, respectively.  The constant $g$ is the gravitational acceleration. $\sigma_i(t)$ is the external disturbance, and $u_i(t)$ is the control input of the $i$th follower vehicle to be designed. The inter-vehicle distance error is given by
\begin{align}\label{be_i}
e_{i}(t) = p_{i-1}(t)-p_i(t)-r_i, \ i=1,2,...,N,
\end{align}
where the constant $r_i$ is the expected inter-vehicle distance. The velocity difference between adjacent vehicles is denoted by $v_{d,i}(t)=v_{i-1}(t)-v_i(t)$. %We have the
\vspace{-1mm}

The control objectives are to design control laws for the follower vehicles so that the following two objectives are satisfied.
\vspace{2mm}
\\
A. string stability: for any given safe inter-vehicle distance error $\delta\in(0,\min_{i=1,2,...,N}r_i)$, there exists a constant $\iota\leq \delta$ and if $\max_{i=1,2,...,N}|e_i(0)|\leq \iota$, then $\max_{i=1,2,...,N}\sup_{t>0}|e_i(t)|\leq \delta$.
\vspace{2mm}\\
B. closed-loop stability: for any given control precision $\epsilon\in(0,\delta]$,  the inter-vehicle distance error $e_i(t)$ satisfies $\limsup_{t\to \infty}|e_i(t)|\leq \epsilon$.
\vspace{-1mm}
\begin{remark}\rm{
Compared with the definition of the string stability in \cite{ploeg2013lp},
the proposed concept of the string stability is concerned with vehicle collision avoidance. From (\ref{be_i}), we know that  $|e_i(t)|\leq \delta$  implies $r_i-\delta\leq p_{i-1}(t)-p_i(t)\leq r_i+\delta, \forall\ t>0$. This together with $\delta<\min_{i=1,2,...,N}r_i$ ensures that the inter-vehicle distance $p_{i-1}(t)-p_i(t)$ is always positive, so vehicle collisions can be avoided.}
\end{remark}
\vspace{-2mm}
\section{Distributed control law based on event-triggered extended state observer}
\vspace{-2mm}
The concept of ESOs was first put forward by \cite{han2009pid}. ESOs do not depend on the accurate system model and are often used to estimate the unmodeled dynamics in the system. Denote $b_i=1/(m_i\tau_i)$ and the vehicle model (\ref{Model}) can be rewritten as
\vspace{-2mm}
\begin{equation}\label{Model1}
\left\{
\begin{aligned}
\dot p_i(t) &= v_i(t), \\
\dot v_i(t) &= a_i(t), \\
\dot a_i(t) &= q_i(t) + \hat b_iu_i(t),\\
\dot q_i(t) &= w_i(t),\ i = 1,2,\ldots,N,
\end{aligned}
\right.
\end{equation}
where
\begin{align}
q_i(t) \hspace{-0.5mm}= &-a_i(t)/\tau_i-c_i v_i^2(t)/(m_i\tau_i)-g\mu_i/\tau_i \nonumber\\
         &-2c_iv_i(t)a_i(t)/m_i+(b_i-\hat b_i)u_i(t)+\sigma_i(t),\label{q_i}\\
w_i(t) \hspace{-0.5mm}=&-\dot a_i(t)/\tau_i-2c_iv_i(t)a_i(t)/(m_i\tau_i)-2c_ia_i^2(t)/m_i\nonumber\\
         &-2c_iv_i(t)\dot a_i(t)/m_i+(b_i-\hat b_i)\dot u_i(t)+\dot \sigma_i(t).\label{w_i}
\end{align}
Here, $q_i(t)$ is the unmodeled dynamics and the constant $\hat b_i$ is the control parameter to be designed.

Next, an ESO is designed to estimate the unmodeled dynamics $q_i(t)$ in the vehicle model (\ref{Model1}). In order to reduce the energy consumption caused by the data transmission, an event-triggered mechanism is embedded in the ESO. The event-triggered ESO is given by
\begin{equation}\label{ESO}
\left\{ %方程组的左边包括大括号\{
\begin{aligned} %设定列阵的格式：{lll} 是三个L，表示三列的对齐方式为Left 对齐
\dot s_i(t) & = -l_is_i(t)-l^2_ia_i(t)-l_i\hat b_i\gamma_i(t), s_i(0)=0, \\
\hat q_i(t) & = s_i(t)+l_ia_i(t), \ i=1,2,...,N,
\end{aligned} %方程列阵的结束\sigma_i(t)
\right. %方程组的右边无符号，利用“.“来标示
\end{equation} %方程组结束
where $\hat q_i(t)$ is the estimate of $q_i(t)$. The constant $l_i$ is the observer gain to be designed. $\gamma_i(t)$ is the input of the ESO, and $s_i(t)$ is a middle variable. Here,
\begin{align}
\gamma_i(t) &= u_i(t_k^i),t\in[t_k^i,t_{k+1}^i),\ k=1,2,...,\label{ET1}\\
t_{k+1}^i &= {\rm inf}\{t>t_k^i|\ |\psi_i(t)|\geq M_i\}, \ t_1^i=0,\label{ET2}
\end{align}
where $t_k^i$ is the $k$th triggering instant of the $i$th follower vehicle, and $\psi_i(t)=\gamma_i(t)-u_i(t)$ is the sampling error. The constant $M_i>0$ is the triggering threshold.

The protocols (\ref{ET1}) and (\ref{ET2}) are the designed event-triggered mechanism. From the triggering instant $t_k^i$ to the next triggering instant $t_{k+1}^i$, the controller stops sending $u_i(t)$ to the ESO. The input of the ESO $\gamma_i(t)$ is unchanged, which is equal to the control input $u_i(t_k^i)$. Until the next triggering instant $t_{k+1}^i$, the controller sends $u_i(t_{k+1}^i)$ to the ESO and the input of the ESO $\gamma_i(t)$ is updated. Due to the event-triggered mechanism, the information transferred from the controller to the ESO is reduced, so the energy consumption caused by data transmission is reduced.

Based on the estimate of the unmodeled dynamics $\hat q_i(t)$, a distributed control law is designed through a modified dynamic surface control method. The dynamic surface control method was first proposed by \cite{swaroop2000dynamic}, which avoids the derivation of the virtual control input by introducing a low-pass filter into each step of the backstepping approach. In order to ensure the string stability, a virtual velocity gain and a virtual acceleration gain are introduced, which can be adjusted to guarantee the string stability.

Before proposing control laws, we need to define some dynamic surfaces. Recall the inter-vehicle distance error $e_i(t)$ and it is treated as the first dynamic surface. The second  dynamic surface is defined as
\begin{align}
z_{1i}(t)=v_i(t)/h_{1i}-\beta_{1i}(t)\label{z_1i},
\end{align}
where the constant $h_{1i}>0$ is the virtual velocity gain, and $\beta_{1i}(t)$ is the output of the first-order low-pass filter
\begin{align}
\kappa_{1i}\dot \beta_{1i}(t)+\beta_{1i}(t)=\alpha_{1i}(t),\ \beta_{1i}(0)=\alpha_{1i}(0)\label{beta_1i},
\end{align}
where the constant $\kappa_{1i}>0$ is filter parameter, and $\alpha_{1i}(t)$ is the first virtual control input to be designed. The third dynamic surface is defined as
\begin{align}
z_{2i}(t)=a_i(t)/h_{2i}-\beta_{2i}(t)\label{z_2i},
\end{align}
where the constant $h_{2i}>0$ is the virtual acceleration gain, and $\beta_{2i}(t)$ is the output of the first-order low-pass filter
\begin{align}
\kappa_{2i}\dot \beta_{2i}(t)+\beta_{2i}(t)=\alpha_{2i}(t),\ \beta_{2i}(0)=\alpha_{1i}(0)\label{beta_2i},
\end{align}
where the constant $\kappa_{2i}>0$ is filter parameter, and $\alpha_{2i}(t)$ is the second virtual control input to be designed. The filtering errors of (\ref{beta_1i}) and (\ref{beta_2i}) are denoted as $\eta_{1i}(t) = \beta_{1i}(t)-\alpha_{1i}(t)$ and $\eta_{2i}(t) = \beta_{2i}(t)-\alpha_{2i}(t)$, respectively. Next, we will show the design process of the control law step by step.

\textbf{Step 1 :}   The design of the first virtual control input $\alpha_{1i}(t)$. By (\ref{be_i}) and (\ref{z_1i}), noting that $\eta_{1i}(t) = \beta_{1i}(t)-\alpha_{1i}(t)$,  we get
\begin{align}\label{de_i}
\dot e_i(t) = &v_{i-1}(t)-v_i(t)\nonumber\\
            = &v_{i-1}(t)-h_{1i}z_{1i}(t)-h_{1i}\eta_{1i}(t)-h_{1i}(t)\alpha_{1i}(t).
\end{align}
The first virtual controller is designed as
\begin{align}\label{alpha_1i}
\alpha_{1i}(t) =  (v_{i-1}(t)+k_{1i}e_{i}(t))/h_{1i}
\end{align}
to stabilize the dynamics (\ref{de_i}), where the constant $k_{1i}>0$ is the controller gain.

\textbf{Step 2 :} The design of the second virtual control input $\alpha_{2i}(t)$. By (\ref{z_1i}) and (\ref{z_2i}), noting that $\eta_{2i}(t) = \beta_{2i}(t)-\alpha_{2i}(t)$,  we get
\begin{align}\label{dz_1i}
\dot z_{1i}(t) = &a_i(t)/h_{1i}-\eta_{1i}(t)/\kappa_{1i}\nonumber\\
            = &h_{2i}(z_{2i}(t)+\eta_{2i}(t)+\alpha_{2i}(t))/h_{1i}-\eta_{1i}(t)/\kappa_{1i},
\end{align}
The second virtual controller is designed as
\begin{align}\label{alpha_2i}
\alpha_{2i}(t) = h_{1i}(-k_{2i}z_{1i}(t)-\eta_{1i}(t)/\kappa_{1i}+h_{1i}e_i(t))/h_{2i}
\end{align}
to stabilize the dynamics (\ref{dz_1i}), where the constant $k_{2i}>0$ is the controller gain.

\textbf{Step 3 :} The design of the control input $u_i(t)$. By (\ref{z_2i}) and (\ref{Model1}), we get
\begin{align}\label{dz_2i}
\dot z_{2i}(t) = &\dot a_i(t)/h_{2i}-\eta_{2i}(t)/\kappa_{2i}\nonumber\\
               = &(q_i(t)+\hat b_iu_i(t))/h_{2i}-\eta_{2i}(t)/\kappa_{2i},
\end{align}
The controller of the $i$th follower vehicle is designed as
\begin{align}\label{u_i}
u_i(t) = &h_{2i}\left(-\hat q_i(t)/h_{2i}-k_{3i}z_{2i}(t)-h_{2i}z_{1i}(t)/h_{1i}\right.\nonumber\\
&\left.-\eta_{2i}(t)/\kappa_{2i}\right)/\hat b_i
\end{align}
to stabilize the dynamics (\ref{dz_2i}), where the constant $k_{3i}>0$ is the controller gain. The block diagram of the designed control law is shown in Fig. 1.
\begin{figure}[!htbp]
	\centering
		\includegraphics[width=8.0cm]{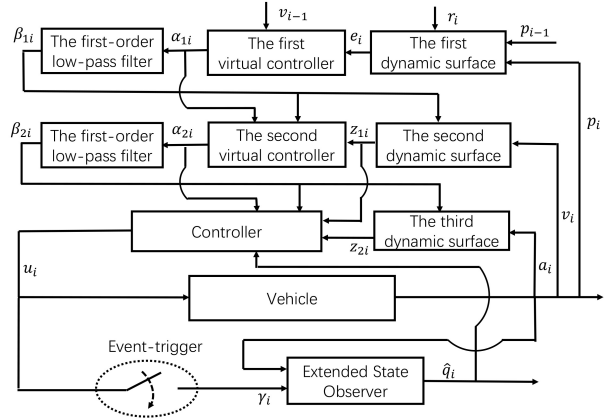}
\caption{ The block diagram of the designed control law. }
\end{figure}

Therefore, the control law of the each follower vehicle consists of the event-triggered ESO (\ref{ESO}), (\ref{ET1}), (\ref{ET2}) and the controller (\ref{u_i}), in which the controller (\ref{u_i}) consists of the virtual controllers (\ref{alpha_1i}), (\ref{alpha_2i}) and the first order low-pass filters (\ref{beta_1i}), (\ref{beta_2i}). It is worth pointing out that the control law of each follower vehicle only uses the information obtained by on-board sensors, such as its own velocity $v_i(t)$, acceleration $a_i(t)$, the velocity of the preceding vehicle $v_{i -1}(t)$ and the inter-vehicle distance $p_{i-1}(t)-p_i(t)$.

\vspace{-2mm}
\section{Stability analysis of vehicle platoon system}
\vspace{-2mm}
We make the following assumptions.

\textbf{A1.} The control input and velocity of the virtual leader vehicle satisfy ${\rm sup}_{t\geq0}|u_0(t)|\leq\overline u_0$ and ${\rm sup}_{t\geq0}|v_0(t)|\leq\overline v_0$, where $\overline u_0>0$ and $\overline v_0>0$ are known constants.

\textbf{A2.} The external disturbance $\sigma_i(t)$ satisfies ${\rm sup}_{t\geq0}|\sigma_i(t)|\leq \sigma_{1i}$. $\sigma_i(t)$ is differentiable and  $\dot \sigma_i(t)$ satisfies ${\rm sup}_{t\geq0} |\dot \sigma_i(t)|\leq \sigma_{2i}$, where $ \sigma_{1i}>0$ and $\sigma_{2i}>0$ are known constants.

\textbf{A3.} The upper and lower bounds of the mass $m_i$, total air resistance coefficient $c_i$, rolling resistance coefficient $\mu_i$ and inertial delay $\tau_i$ of the $i$th follower vehicle are known, which are denoted by $\overline m_i, \ \underline m_i,\ \overline c_i,\ \underline c_i,\ \overline \mu_i,\ \underline \mu_i,\ \overline \tau_i,\ \underline \tau_i$, respectively.

The observation error of the event-triggered ESO is denoted by $e_{1i}(t)=q_i(t)-\hat q_i(t)$. For the boundedness of the observation errors, we have the following lemma.
\vspace{-2mm}
\begin{lemma}\rm{
Suppose that Assumptions A2$-$A3 hold. If the differentiation of the unmodeled dynamics $\dot q_i(t)$ satisfies $|\dot q_i(t)|\leq c_{1i}+c_{2i}|e_{1i}(t)|+\overline c_{bi}l_i|e_{1i}(t)|+\overline c_{bi}l_i\hat b_i|\psi_i(t)|$, the estimation of control gain $\hat b_i$ satisfies $\max\left\{\underline b_i,\overline b_i/2\right\}<\hat b_i\leq\overline b_i$,  the observer gain $l_i$ satisfies $l_i>(c_{1i}+c_{2i}\overline e_{1i})/((1-\overline c_{bi})\overline e_{1i})$ and the triggering threshold $M_i=\overline e_{1i}(l_i-c_{2i}-\overline c_{bi}l_i)-c_{1i})/(l_i\hat b_i(1+\overline c_{bi}))$, then ${\rm sup}_{t\geq0}|e_{1i}(t)|\leq\overline e_{1i}$, where $c_{1i}$ and $c_{2i}$ are positive constants, $\overline e_{1i} =|a_i(0)|/\tau_i+\overline c_i v_i^2(0)/(\underline m_i\underline \tau_i)+g\overline \mu_i/\underline \tau_i+\max\{\overline b_i-\hat b_i,\hat b_i-\underline b_i\}|u_i(0)|+\sigma_{1i}+|\hat q_i(0)|$, $\overline c_{bi}=\max\left\{(\overline b_i-\hat b_i)/\hat b_i,(\hat b_i-\underline b_i)/\hat b_i\right\}$, $\overline b_i = 1/(\underline m_i\underline \tau_i)$, $\underline b_i = 1/(\overline m_i\overline \tau_i)$}.
\end{lemma}
The proof of Lemma 1 is given in Appendix A.

For stability of the vehicle platoon system, we have the following theorem.
\vspace{-2mm}
\begin{thm}\rm{
 Suppose that Assumptions A1$-$A3 hold. Consider the system (\ref{Model0}) and (\ref{Model}) under the distributed control law which consists of the event-triggered ESO (\ref{ESO}), (\ref{ET1}), (\ref{ET2}) and the controller (\ref{u_i}). For any given safe inter-vehicle distance error $\delta\in(0,\min_{i=1,2,...,N} r_i)$ and control precision $\epsilon\in(0,\delta]$, if the control parameters are designed according to Conditions C1$-$C3, then the string stability and closed-loop stability of the vehicle platoon system are guaranteed, and the Zeno behavior is avoided.}

\vspace{-2mm}
C1. The virtual velocity gain $h_{1i}$ and the virtual acceleration gain $h_{2i}$, $i=1,2,...,N$ satisfy $v_{d,i}^2(0)/h_{1i}^2+(a_i(0)-k_{2i}v_{d,i}(0))^2/h_{2i}^2\leq \delta^2$.

C2. The controller gains $k_{1i}$, $k_{2i}$, $k_{3i}$, $i=1,2,...,N$ satisfy
\vspace{-2mm}
\begin{equation}\label{parameters1}
\left\{
\begin{aligned}
&k_{1i}\geq (3\xi_i+\epsilon^2)/(2\epsilon^2),\\
&k_{2i}\geq (3\xi_i+\epsilon^2)/(2\epsilon^2),\\
&k_{3i}\geq (3\xi_i^2h_{2i}^2+\overline e_{1i}^2\epsilon^2)/(2h_{2i}^2\epsilon^2\xi_i),
\end{aligned}
\right.
\end{equation}
where $\xi_i$ is an arbitrarily given positive constant, $i=1,2,...,N$.

C3. The low-pass filter parameters $\kappa_{1i}$, $\kappa_{2i}$, $i=1,2,...,N$ satisfy
\begin{equation}\label{parameters2}
\left\{
\begin{aligned}
&\kappa_{1i}\leq (2\xi_i\epsilon^2)/(3\xi_i^2+\xi_i\epsilon^2h_{1i}^2+\epsilon^2\alpha^2_{3i}),\\
&\kappa_{2i}\leq (2\xi_i\epsilon^2h_{1i}^2)/(3\xi_i^2h_{1i}^2+\xi_i\epsilon^2h_{2i}^2+h_{1i}^2\epsilon^2\alpha^2_{4i}),
\end{aligned}
\right.
\end{equation}
where  $\alpha_{3i} = \overline a_{i-1}/h_{1i}+\left(2k_{1i}+k_{1i}^2/h_{1i}\right)\delta$,  $\alpha_{4i} =[(k_{1i}h_{1i}^2+k_{2i}h_{1i}^2+|h_{1i}/\kappa_{1i}^2-h_{1i}^3|+|h_{1i}k_{2i}^2-h_{1i}^3|)/h_{2i}
 +\alpha_{3i}/\kappa_{1i}+2k_{2i}]\delta$, $\overline v_i =  (2h_{1i}+k_{1i})\delta+\overline v_{i-1}$, $\overline a_0 = \max\{a_i(0), \overline u_0\}$, $\overline a_i =  (2h_{2i}+h_{1i}k_{2i}+h_{1i}^2+h_{1i}/\kappa_{1i})\delta$.
\end{thm}
\vspace{-2mm}
The proof of Theorem 1 is given in Appendix B. From Theorem 1, it is shown that the control parameters can be
properly designed to ensure the stability of the vehicle platoon system provided the disturbance to the leader vehicle $u_0(t)$ is bounded.

\vspace{-2mm}
\section{Numerical simulations}
\vspace{-2mm}
Firstly, the vehicle models (\ref{Model0}) and (\ref{Model}) are considered. Suppose that there is one virtual leader vehicle and eight follower vehicles in the platoon.
The upper and lower bounds of the mass $m_i$, inertial delay $\tau_i$, total air resistance coefficient $c_i$ and rolling resistance coefficient $\mu_i$ of the follower vehicles are given by $\overline m_i=2000\, \rm{kg}$,\ $\underline m_i=1500\, \rm{kg}$,\ $\overline \tau_i=0.4$, \ $\underline \tau_i=0.2$,\ $\overline c_i=0.4$,\ $\underline c_i=0.2$,\ $\overline \mu_i=0.05$,\ $\underline \mu_i=0.02$,\ $i=1,2,...,8$, respectively. The model parameters of the follower vehicles are randomly generated within the upper and lower bounds given above.

The expected inter-vehicle distances between adjacent vehicles and safe inter-vehicle distance errors are given by  $r_i=8\, \rm{m}$, \, $i=1,2,...,8$ and $\delta=7\, \rm{m} $, respectively. The external disturbances are taken as $\sigma_i(t)=\lambda_{1i}e^{-\lambda_{2i}t}+\lambda_{3i}\sin(\lambda_{4i}t)$, where the constants $\lambda_{1i},\lambda_{2i},\lambda_{3i}$ and $\lambda_{4i}$ are randomly generated in $[1\, ,20],\ [0.1\, ,0.5],\ [0.5\, ,1]$ and $[4\, ,8]$, respectively. The initial position, velocity and acceleration of the virtual leader vehicle are given by $p_0(0)=80\, \rm{m}$, $v_0(0)=10\, \rm{m/s}$ and $a_0(0)=0\, \rm{m/s^2}$, respectively. The initial positions, velocities and accelerations of the follower vehicles are given in Table 1.
\begin{table}[H]
\setlength\tabcolsep{4pt}
\caption{The initial positions, velocities and accelerations of the follower vehicles.}
\begin{tabular}{ccccccccc}
   \toprule
   i & 1 & 2 & 3 & 4 & 5 & 6 & 7 & 8  \\
   \midrule
   $p_i(0)(\rm{m})$ & 71 & 63.5 & 54 & 47.2 & 38.4 & 30.6 & 22.1 & 14.8 \\
   $v_i(0)(\rm{m/s})$ & 10 & 11 & 11.5 & 12.5 & 12.5 & 11.5 & 13.5 & 13 \\
   $a_i(0)(\rm{m/s^2})$ & 0 & 1.5 & -1 & 0 & -2 & 1 & 0 & -1\\
   \bottomrule
\end{tabular}
\end{table}
The maneuver of the virtual leader vehicle is divided into three stages. \textbf{Stage 1} ($t\in[0\,\rm{s},\,6\,s)$): the virtual leader vehicle keeps moving at a constant velocity. \textbf{Stage 2} ($t\in[6\,\rm{s},\,9\,\rm{s})$): the virtual leader vehicle accelerates, and the expected acceleration of the leader vehicle is generated by CarSim. Specifically,  we construct a virtual vehicle in CarSim and then control its throttle opening to accelerate, and record the acceleration. Then we regard the recorded acceleration as the expected acceleration of the virtual leader vehicle. \textbf{Stage 3} ($t\in[9\,\rm{s},\,15\,\rm{s})$): the virtual leader vehicle continues to move at a constant velocity after accelerating.

Before carrying out the simulations with the designed controller (\ref{u_i}),
we will show the vehicle platoon maneuver with controller
\begin{align}\label{su_i}
u_i(t)=k_pe_i(t)+k_vv_{d,i}(t)+k_aa_{i-1}(t)+k_da_i(t),
\end{align}
which performs well for the vehicle with linear dynamics \citep{Rajamani2002Semi}. We choose $k_p=2000,\ k_v=4000,\ k_a=2000,\ k_d=100$. The evolution of the inter-vehicle distance errors with the controller (\ref{su_i}) is shown in Fig. 2.
\begin{figure}[H]
		\centering
		\includegraphics[width=7.5cm]{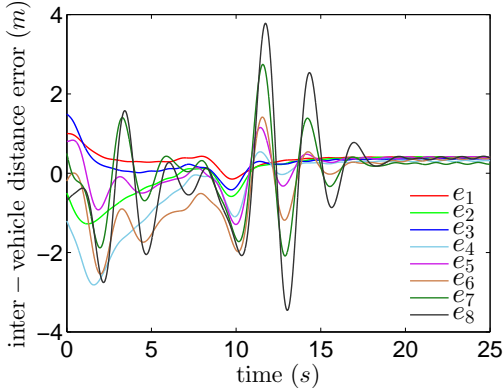}
\caption{ Inter-vehicle distance errors with the controller (\ref{su_i}). }
\end{figure}
Then we preform the simulations with the designed
controller (19). For $\epsilon=0.1$, we choose $k_{1i}=0.8,\ k_{2i}=1.5,\ k_{3i}=300,\ \kappa_{1i}=0.05,\ \kappa_{2i}=0.01,\ l_i=1200,\ \hat b_i=0.003,\ h_{1i}=2,\ h_{2i}=8,\ i=1,2,...,8$.
The evolution of the inter-vehicle distance errors with $\epsilon=0.1$ is shown in Fig. 3. The actual and the estimated unmodeled dynamics of the 1st and 7th follower vehicles are shown in Fig. 4. The triggering instants of event-triggered ESOs are shown in Fig. 5.
\begin{figure}[H]
		\centering
		\includegraphics[width=7.5cm]{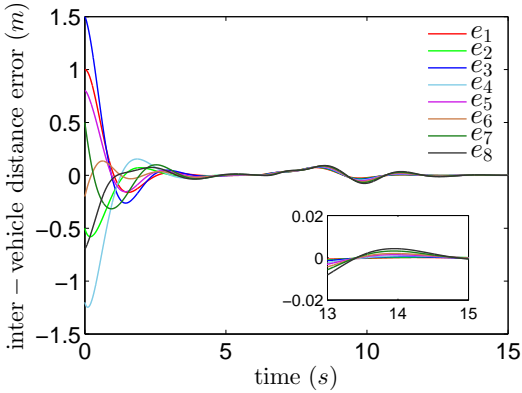}
\caption{ Inter-vehicle distance errors ($\epsilon=0.1$). }
\end{figure}
\begin{figure}
		\centering
		\includegraphics[width=7.5cm]{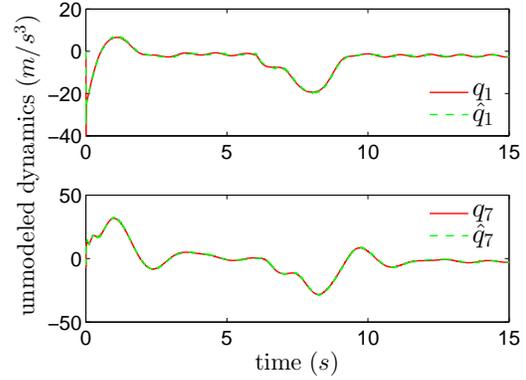}
\caption{ The actual and the estimated unmodeled dynamics of the 1st and 7th follower vehicles.}
\end{figure}
\begin{figure}
		\centering
		\includegraphics[width=8.5cm]{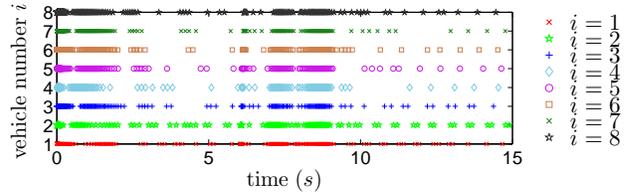}
\caption{ The triggering instants of event-triggered ESOs.}
\end{figure}
\vspace{-2mm}
Comparing Fig. 2 and Fig. 3, it can be seen that the controller (\ref{u_i}) designed based on the nonlinear vehicle model has better performance than the controller (\ref{su_i}).
From Fig. 3, it is shown that the inter-vehicle errors gradually decrease from the initial non-zero states and finally converge to a small neighborhood of zero, then the inter-vehicle errors deviate from the steady states due to the acceleration of the leader vehicle. As the acceleration of the leader vehicle becomes zero, the inter-vehicle distance errors converge again. The absolute values of the inter-vehicle distance error $|e_i(t)|$ never exceeds the given safe value $\delta$. Fig. 3 shows that the designed control laws can deal with not only the non-zero initial states but also the disturbances to the leader vehicle.
It can be seen from Fig. 4 that the unmodeled dynamics in different vehicle models can be estimated well by the event-triggered ESOs.  Fig. 5 shows that the event-triggered mechanism can effectively reduce the data transmission from the controller to the ESO.

Then for $\epsilon=0.01$, we choose $k_{1i}=2,\ k_{2i}=8,\ k_{3i}=1000,\ \kappa_{1i}=0.005,\ \kappa_{2i}=0.001,\ l_i=1200,\ \hat b_i=0.003,\ h_{1i}=2,\ h_{2i}=8,\ i=1,2,...,8$. The evolution of the inter-vehicle distance errors with $\epsilon=0.01$ is shown in Fig. 6.
\begin{figure}
		\centering
		\includegraphics[width=7.5cm]{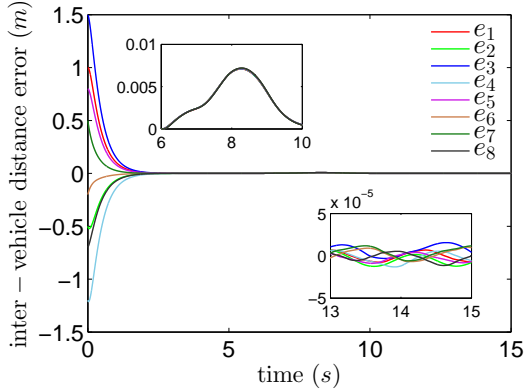}
\caption{ Inter-vehicle distance errors ($\epsilon=0.01$). }
\end{figure}
It can be seen from Fig. 3 and Fig. 6 that the control precision can be adjusted by properly choosing the control parameters. From Fig. 3 and Fig. 6, we know that the larger
controller gains $k_{1i}$, $k_{2i}$, $k_{3i}$ and the smaller filter parameters $\kappa_{1i}$, $\kappa_{2i}$ lead to a faster convergence and reduce the fluctuations of the inter-vehicle distance errors caused by the disturbances to the virtual leader vehicle. However, the larger controller gains may result in larger control inputs when the initial inter-vehicle distance errors are large, which may leads to input saturation, so we need to choose the control parameters with a compromise.

Next, the vehicle models in CarSim are considered and the joint simulations of Simulink and CarSim are carried out. Compared with the nonlinear vehicle models obtained by performing a series of simplifications on the dynamic characteristics of real vehicles, the vehicle models in CarSim are more complex and more consistent with the dynamic characteristics of real vehicles. Firstly, one leader vehicle model and eight follower vehicle models are constructed in CarSim, which are added to the model library of Simulink in the form of S-functions. Then, the joint simulation platform of Simulink and CarSim is built. The constructed vehicle models in CarSim are called by Simulink directly. Similar to the previous simulation, the maneuver of the leader vehicle is divided into three stages. First, the leader vehicle moves at a constant velocity, then slows down, and finally moves at a constant velocity after decelerating.
In order to facilitate testing, we develop a graphical user interface (GUI) based on MATLAB. The control parameters can be edited in the GUI directly and the inter-vehicle distance errors can be automatically displayed in the GUI after the program runs. We reselect $k_{1i}=0.9,\ k_{2i}=20,\ k_{3i}=100,\ \kappa_{1i}=0.05,\ \kappa_{2i}=0.005,\ l_i=800,\ \hat b_i=0.05,\ h_{1i}=2,\ h_{2i} = 4,\ i=1,2,...,8$. The GUI based on MATLAB is shown in Fig. 7.
\begin{figure}[!htbp]
	\centering
		\includegraphics[width=8.2cm]{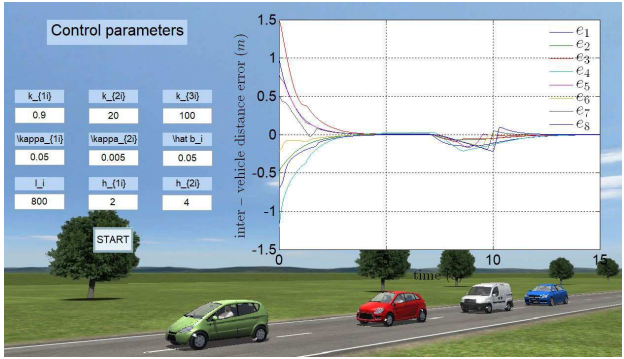}
\label{Fig1}
\caption{ The GUI based on MATLAB. }
\end{figure}

It can be seen from Fig. 7 that the inter-vehicle distance errors stay within a safe range and eventually converge to a small neighborhood of zero, either with non-zero initial states or with disturbances to the leader vehicle. The joint simulation of Simulink and CarSim shows that although the control laws in this paper are designed for the simplified vehicle models, they are still effective for the more complicated vehicle models in CarSim.
\vspace{-2mm}
\section{Conclusion}
\vspace{-2mm}
We have presented a distributed control law based on an event-triggered ESO and a modified dynamic surface control method for the platoon control of the third-order nonlinear vehicle. The constant spacing policy has been adopted, and the vehicle model with parameter uncertainties and external disturbances has been considered. The control law of each follower vehicle only uses its own velocity, acceleration, the velocity of preceding vehicle and the inter-vehicle distance, which can be obtained by on-board sensors. Firstly, an event-triggered ESO has been designed to estimate the unmodeled dynamics. Then based on the estimate of the unmodeled dynamics, a distributed control law has been proposed by utilizing a modified dynamic surface control method. We have analyzed the stability of the vehicle platoon system and given the range of the control parameters to ensure that the observation errors of the ESOs are bounded and that the string stability and closed-loop stability are guaranteed. We have proved that the Zeno behavior is avoided under the designed event-triggered mechanism.

In this paper, the input delay and the input saturation are not considered. In the future, it is challenging to consider how to design a control law that can deal with the input delay and the input saturation simultaneously.

\vspace{-2mm}
\section*{Acknowledgements}
\vspace{-2mm}
This work was supported in part by the National Natural Science Foundation of China under Grant 61977024 and in part by the Basic
Research Project of Shanghai Science and Technology Commission under
Grant 20JC1414000.
\vspace{-2mm}
\begin{appendices}
\section{Proof of Lemma 1}
\textbf{Proof of Lemma 1.}
By (\ref{Model1}), (\ref{ESO}) and (\ref{ET1}), we know
\begin{align}\label{dhat_q}
\dot {\hat q}_i(t) =& \dot s_i(t)+l_i\dot a_i(t)\nonumber \\
=& -l_is_i(t)-l^2_ia_i(t)-l_i\hat b_i\gamma_i(t)+l_i(q_i(t)+\hat b_iu_i(t))\nonumber \\
%=& -l_i(\hat q_i(t)-l_ia_i(t))-l^2_ia_i(t)-l_i\hat b_i\gamma_i(t)+l_i(q_i(t)+\hat b_iu_i(t))\nonumber \\
=&  l_ie_{1i}(t)-l_i\hat b_i\psi_i(t).
\end{align}
By (\ref{Model1}) and (\ref{dhat_q}), we have the following observation error dynamics
\begin{align}\label{Psi}
\dot e_{1i}(t)= -l_ie_{1i}(t)+l_i\hat b_i\psi_i(t)+w_i(t).
\end{align}
Consider the Lyapunov function $V_{ei}(t)=e^2_{1i}(t)/2$. By (\ref{Psi}), noting that $\left|w_i(t)\right|\leq c_{1i}+c_{2i}\left|e_{1i}(t)\right|+\overline c_{bi}l_i\left|e_{1i}(t)\right|+\overline c_{bi}l_i\hat b_i\left|\psi_i(t)\right|$, we get
\begin{align}\label{detaV}
\dot V_{ei}(t)
%= & e_{1i}(t)\dot e_{1i}(t)\nonumber \\
              = & -l_ie^2_{1i}(t)+l_i\hat b_i\psi_i(t)e_{1i}(t)+e_{1i}(t)w_i(t)\nonumber \\
            \leq& -l_ie^2_{1i}(t)+l_i\hat b_i\left|\psi_i(t)\right|\left|e_{1i}(t)\right|+\left|e_{1i}(t)\right|\left|w_i(t)\right| \nonumber \\
           %\leq & -l_ie^2_{1i}(t)+\left|e_{1i}(t)\right|\left(l_i\hat b_i\left|\psi_i(t)\right|+c_{1i}+c_{2i}\left|e_{1i}(t)\right|+\overline c_{bi}l_i\left|e_{1i}(t)\right|+\overline c_{bi}l_i\hat b_i\left|\psi_i(t)\right|\right)   \nonumber \\
             \leq &  -l_ie^2_{1i}(t)+c_{2i}e^2_{1i}(t)+\overline c_{bi}l_ie^2_{1i}(t)+\left[c_{1i}\right.\nonumber\\
             &\left.+l_i\hat b_i(1+\overline c_{bi})\left|\psi_i(t)\right|\right]\left|e_{1i}(t)\right|\nonumber \\
              = & -2(l_i-c_{2i}-\overline c_{bi}l_i)V_{ei}(t)+\sqrt{2}\left[c_{1i}+l_i\hat b_i(1\right.\nonumber\\
              &\left.+\overline c_{bi})\left|\psi_i(t)\right|\right]\sqrt{V_{ei}(t)}.
\end{align}
By (\ref{detaV}), noting that $\frac{d V_{ei}(t)}{dt}=2\sqrt{V_{ei}(t)}\frac{d\sqrt{ V_{ei}(t)}}{dt}$, we obtain
\begin{align}\label{deta11}
\sqrt{V_{ei}(t)}\leq&\left(\sqrt{V_{ei}(0)}-A_i\right)e^{-\left(l_i-c_{2i}-\overline c_{bi}l_i\right)t}+ A_i,
\end{align}
where$A_i=\sqrt{2}[c_{1i}+l_i\hat b_i(1+\overline c_{bi})|\psi_i(t)|]/[2(l_i-c_{2i}-\overline c_{bi}l_i)]$.
By (\ref{deta11}), noting that $V_{ei}(t)=e_{1i}^2(t)/2$, we obtain
\begin{align}\label{e_i}
&|e_{1i}(t)|\nonumber \\
=&\sqrt{2V_{ei}(t)}\nonumber \\
\leq&\left[\sqrt{2V_{ei}(0)}-\sqrt{2}A_i\right]e^{-\left(l_i-c_{2i}-\overline c_{bi}l_i\right)t}+\sqrt{2}A_i\nonumber \\
%= & \left[|e_{1i}(0)|-\frac{c_{1i}+l_i\hat b_i(1+\overline c_{bi})\left|\psi_i(t)\right|}{l_i-c_{2i}-\overline c_{bi}l_i }\right]e^{-\left(l_i-c_{2i}-\overline c_{bi}l_i\right)t}+\frac{c_{1i}+l_i\hat b_i(1+\overline c_{bi})\left|\psi_i(t)\right|}{l_i-c_{2i}-\overline c_{bi}l_i}.
= & \left(|e_{1i}(0)|-\sqrt{2}A_i\right)e^{-\left(l_i-c_{2i}-\overline c_{bi}l_i\right)t}+\sqrt{2}A_i.
\end{align}
By $\max\left\{\underline b_i,\overline b_i/2\right\}<\hat b_i\leq\overline b_i$, we know $\overline c_{bi}<1$. According to  $l_i>(c_{1i}+c_{2i}\overline e_{1i})/[(1-\overline c_{bi})\overline e_{1i}]$ and $\overline c_{bi}<1$, we have $\overline e_{1i}(l_i-c_{2i}-\overline c_{bi}l_i)-c_{1i}>0$. By (\ref{ET1}) and  (\ref{ET2}), noting that  $M_i=\overline e_{1i}(l_i-c_{2i}-\overline c_{bi}l_i)-c_{1i})/(l_i\hat b_i(1+\overline c_{bi}))$, we get $|\psi_i(t)|<[\overline e_{1i}(l_i-c_{2i}-\overline c_{bi}l_i)-c_{1i}]/(l_i\hat b_i(1+\overline c_{bi})),\ t\in[t_k^i, t_{k+1}^i)$. This together with (\ref{e_i}) leads to
\begin{align}\label{e_i1}
|e_{1i}(t)|
\leq  \left(|e_{1i}(0)|-\overline e_{1i}\right)e^{-\left(l_i-c_{2i}-\overline c_{bi}l_i\right)t}+\overline e_{1i}.
\end{align}

From Assumption A2, we get ${\rm sup}_{t\geq0}|\sigma_i(t)|\leq \sigma_{1i}$. By (\ref{q_i}), we get $e_{1i}(0)=-a_i(0)/\tau_i-c_i v_i^2(0)/(m_i\tau_i)-g\mu_i/\tau_i
+(b_i-\hat b_i)u_i(0)+\sigma_i(0)-\hat q_i(0)$.
This together with  ${\rm sup}_{t\geq0}|\sigma_i(t)|\leq \sigma_{1i}$ and Assumption A3 leads to $|e_{1i}(0)|\leq\overline e_{1i}$. By (\ref{e_i1}), noting that $|e_{1i}(0)|\leq\overline e_{1i}$, we have $|e_{1i}(t)|\leq\overline e_{1i},\ \forall\ t\geq0$. \qed

\section{Proof of Theorem 1}
\textbf{Proof of Theorem 1.} 	
Choose the following Lyapunov function $V_i(t) = (e_i^2(t)+z_{1i}^2(t)+z_{2i}^2(t)+\eta_{1i}^2(t)+\eta_{2i}^2(t))/2$.

By (\ref{be_i}), (\ref{Model1}), (\ref{alpha_1i})$-$(\ref{u_i}), the time derivative of $V_i(t)$ is calculated as
\begin{align}\label{dV_i}
&\dot V_i(t) \nonumber\\
           =&-k_{1i}e^2_i(t)-k_{2i}z^2_{1i}(t)-k_{3i}z_{2i}^2(t)-\eta_{1i}^2(t)/\kappa_{1i}\nonumber\\
            &-\eta_{2i}^2(t)/\kappa_{2i}-h_{1i}e_i(t)\eta_{1i}(t)+h_{2i}z_{1i}(t)\eta_{2i}(t)/h_{1i}\nonumber\\
            &+z_{2i}(t)e_{1i}(t)/h_{2i}-\eta_{1i}(t)\dot \alpha_{1i}(t)-\eta_{2i}(t)\dot \alpha_{2i}(t) \nonumber\\
       \leq& -k_{1i}e^2_i(t)-k_{2i}z^2_{1i}(t)-k_{3i}z_{2i}^2(t)-\eta^2_{1i}(t)/\kappa_{1i}\nonumber\\
           &-\eta^2_{2i}(t)/\kappa_{2i}+h_{1i}|e_i(t)||\eta_{1i}(t)|+h_{2i}|z_{1i}(t)||\eta_{2i}(t)|/h_{1i}\nonumber\\
           &+|z_{2i}(t)||e_{1i}(t)|/h_{2i}+|\eta_{1i}(t)||\dot \alpha_{1i}(t)|+|\eta_{2i}(t)||\dot \alpha_{2i}(t)| \nonumber\\
       \leq& -k_{1i}e^2_i(t)-k_{2i}z^2_{1i}(t)-k_{3i}z_{2i}^2(t)-\eta^2_{1i}(t)/\kappa_{1i}\nonumber\\
           &-\eta^2_{2i}(t)/\kappa_{2i}+ e_i^2(t)/2+h_{1i}^2\eta_{1i}^2(t)/2+z_{1i}^2(t)/2\nonumber\\
           &+h_{2i}^2\eta_{2i}^2(t)/(2h_{1i}^2)+z_{2i}^2(t)e_{1i}^2(t)/(2h_{2i}^2\xi_i)\nonumber\\
           &+\eta_{1i}^2(t)\dot \alpha_{1i}^2(t)/(2\xi_i)+\eta_{2i}^2(t)\dot \alpha_{2i}^2(t)/(2\xi_i)+3\xi_i/2 \nonumber\\
          =&-\left(k_{1i}-1/2\right)e^2_i(t)-\left(k_{2i}-1/2\right)z^2_{1i}(t)\nonumber\\
           &-\left[k_{3i}-e_{1i}^2(t)/(2h_{2i}^2\xi_i)\right]z_{2i}^2(t)\nonumber\\
           &-\left[1/\kappa_{1i}-h_{1i}^2/2-\dot \alpha^2_{1i}(t)/(2\xi_i)\right]\eta^2_{1i}(t)\nonumber\\
&-\left[1/\kappa_{2i}-h_{2i}^2/(2h_{1i}^2)-\dot \alpha_{2i}^2(t)/(2\xi_i)\right]\eta_{2i}^2(t)+3\xi_i/2.
\end{align}
Consider the  compact set
\begin{align*}
\Omega:=\{(x_{1},x_{2},x_{3},x_{4},x_{5}):&x^2_{1}/2+x^2_{2}/2+x^2_{3}/2\\
                 &+x^2_{4}/2+x^2_{5}/2\leq \delta^2/2\}.
\end{align*}
By (\ref{z_1i}) and (\ref{alpha_1i}) , we obtain
\begin{align}\label{v_i}
v_i(t)=&h_{1i}(z_{1i}(t)+\beta_{1i}(t))\nonumber\\
      %=&h_iz_{1i}(t)+h_i\eta_{1i}(t)+h_i\alpha_{1i}(t)\nonumber\\
      =&h_{1i}z_{1i}(t)+h_{1i}\eta_{1i}(t)+v_{i-1}(t)+k_{1i}e_i(t).
\end{align}
From (\ref{z_2i}) and (\ref{alpha_2i}), we know
\begin{align}\label{a_i}
a_i(t)=&h_{2i}(z_{2i}(t)+\beta_{2i}(t))\nonumber\\
      %=&z_{2i}(t)+\eta_{2i}(t)+\alpha_{2i}(t)\nonumber\\
      =&h_{2i}z_{2i}(t)+h_{2i}\eta_{2i}(t)-h_{1i}k_{2i}z_{1i}(t)\nonumber\\
       &-h_{1i}\eta_{1i}(t)/\kappa_{1i}+h_{1i}^2e_i(t).
\end{align}
By (\ref{be_i}), (\ref{z_1i}) and (\ref{alpha_1i}), we have
\begin{align} \label{dalpha_1i}
\dot \alpha_{1i}(t)
=&(\dot v_{i-1}(t)+k_{1i}\dot e_i(t))/h_{1i}\nonumber\\
                =&a_{i-1}(t)/h_{1i}-k_{1i}z_{1i}(t)-k_{1i}\eta_{1i}(t)\nonumber\\
                &-k_{1i}^2e_i(t)/h_{1i}.
\end{align}
By (\ref{z_1i}), (\ref{beta_2i}) and (\ref{alpha_2i}), we know
\begin{align} \label{dalpha_2i}
\dot \alpha_{2i}(t)
=&h_{1i}(-k_{2i}\dot z_{1i}(t)-\dot \eta_{1i}(t)/\kappa_{1i}+h_{1i}\dot e_i(t))/h_{2i}\nonumber\\
=&-\left(h_{1i}^2k_{1i}+h_{1i}^2k_{2i}\right)e_i(t)/h_{2i}+\dot \alpha_{1i}(t)/\kappa_{1i}\nonumber\\
 &+ \left(h_{1i}/\kappa_{1i}^2-h_{1i}^3\right)\eta_{1i}(t)/h_{2i}-k_{2i}\eta_{2i}(t)\nonumber\\
&+\left(h_{1i}k_{2i}^2-h_{1i}^3\right)z_{1i}(t)/h_{2i}-k_{2i}z_{2i}(t).
\end{align}
According to (\ref{Model1}) and (\ref{u_i}), we obtain
\begin{align} \label{da_i}
&\dot a_i(t) \nonumber \\
= &q_i(t)+\hat b_iu_i(t)\nonumber \\
            = &e_{1i}(t)-(k_{3i}z_{2i}(t)+h_{2i}z_{1i}(t)/h_{1i}+\eta_{2i}(t)/\kappa_{2i})h_{2i}.
\end{align}
By (\ref{v_i}) and Assumption A1, we know that $v_i(t)$ satisfies $|v_i(t)|\leq \overline v_i,\ \forall\ t\geq0$ on the compact set $\Omega$. By (\ref{a_i}), we know that $a_i(t)$ satisfies $|a_i(t)|\leq \overline a_i,\ \forall\ t\geq0$ on the compact set $\Omega$. By (\ref{dalpha_1i}), (\ref{dalpha_2i}) and Assumption A1 and noting that $|a_i(t)|\leq \overline a_i,\ \forall\ t\geq0$, we obtain that $\dot \alpha_{1i}(t)$ and $\dot \alpha_{2i}(t)$ satisfy $|\dot \alpha_{1i}(t)|\leq \alpha_{3i}$ and $|\dot \alpha_{2i}(t)|\leq\alpha_{4i},\ \forall\ t\geq0$ on the compact set $\Omega$, respectively. From (\ref{da_i}), we know that $\dot a_i(t)$ satisfies $|\dot a_i(t)| \leq |e_{1i}(t)|+\left(k_{3i}+h_{2i}/h_{1i}+1/\kappa_{2i}\right)h_{2i}\delta$ on the compact set $\Omega$.

By (\ref{Model1}), (\ref{beta_1i})$-$(\ref{u_i}) and (\ref{dhat_q}), we know $\dot u_i(t) = h_{2i}[-(l_i+k_{3i})e_{1i}(t)/h_{2i}+l_i\hat b_i\psi_i(t)/h_{2i}-h_{2i}e_i(t)+(k_{2i}+k_{3i})h_{2i}z_{1i}(t)/h_{1i}+(k_{3i}^2-h_{2i}^2/h_{1i}^2)z_{2i}(t)-(h_{2i}^2/h_{1i}^2-1/\kappa_{2i}^{2})\eta_{2i}(t)+\dot \alpha_{2i}(t)/\kappa_{2i}]/\hat b_i$.
This together with $|\dot \alpha_{2i}(t)|\leq\alpha_{4i},\ \forall\ t\geq0$ leads to
\begin{align} \label{u_i2}
&|\dot u_i(t)|\nonumber\\
\leq &h_{2i}\left[(l_i+k_{3i})|e_{1i}(t)|/h_{2i}+l_i\hat b_i|\psi_i(t)|/h_{2i}+h_{2i}\delta \right.\nonumber\\
&\left.+(k_{2i}+k_{3i})h_{2i}\delta/h_{1i}+\left|k_{3i}^2-h_{2i}^2/h_{1i}^2\right|\delta\right.\nonumber\\
&\left.+\left|h_{2i}^2/h_{1i}^2-1/\kappa_{2i}^{2}\right|\delta+\alpha_{4i}/\kappa_{2i} \right]/\hat b_i.
\end{align}
From Assumption A2, we know ${\rm sup}_{t\geq0} |\dot \sigma_i(t)|\leq  \sigma_{2i}$. By (\ref{w_i}), (\ref{u_i2}) and Assumption A3, noting that ${\rm sup}_{t\geq0} |\dot \sigma_i(t)|\leq  \sigma_{2i}$, $|v_i(t)|\leq \overline v_i$, $|a_i(t)|\leq \overline a_i$ and $|\dot a_i(t)| \leq |e_{1i}(t)|+\left(k_{3i}+h_{2i}/h_{1i}+1/\kappa_{2i}\right)h_{2i}\delta$, we know
\begin{align} \label{absw_i}
&|w_i(t)|\nonumber\\
\leq& |\dot a_i(t)|/\tau_i+2 c_i|v_i(t)||a_i(t)|/(m_i \tau_i)+2 c_i|v_i(t)||\dot a_i(t)|\nonumber\\
&/m_i+2 c_ia_i^2(t)/ m_i+|(b_i-\hat b_i)/\hat b_i||\dot u_i(t)|+|\dot \sigma_i(t)|\nonumber\\
\leq& \left(1/\underline \tau_i+2\overline c_i\overline v_i/\underline
  m_i\right)\left[|e_{1i}(t)|+\left(k_{3i}+h_{2i}/h_{1i}+1/\kappa_{2i}\right)\right.\nonumber\\
  &\left.\times h_{2i}\delta\right]+2\overline c_i\overline v_i\overline a_i/(\underline m_i\underline \tau_i)+2\overline c_i\overline a_i^2/\underline m_i+\sigma_{2i}\nonumber\\
  &+|(b_i-\hat b_i)/\hat b_i|\left[(l_i+k_{3i})|e_{1i}(t)|+l_i\hat b_i|\psi_i(t)|+h_{2i}^2\delta\right.\nonumber\\
  &\left.+(k_{2i}+k_{3i})h_{2i}^2\delta/h_{1i}+|k_{3i}^2-h_{2i}^2/h_{1i}^2|\delta+\alpha_{4i}/\kappa_{2i}\right.\nonumber\\
  &\left.+\left|h_{2i}^2/h_{1i}^2-1/\kappa_{2i}^{2}\right|\delta\right].
\end{align}
Denote $c_{bi}=(b_i-\hat b_i)/\hat b_i$. By $\max\left\{\underline b_i,\overline b_i/2\right\}<\hat b_i\leq\overline b_i$, we know $|c_{bi}|\leq\overline c_{bi}$ and $\overline c_{bi}<1$. From (\ref{absw_i}), noting that $|\dot a_i(t)| \leq |e_{1i}(t)|+\left(k_{3i}+h_{2i}/h_{1i}+1/\kappa_{2i}\right)h_{2i}\delta$, we obtain
\begin{align*}
\left|w_i(t)\right|\leq c_{1i}+c_{2i}\left|e_{1i}(t)\right|+\overline c_{bi}l_i\left|e_{1i}(t)\right|+\overline c_{bi}l_i\hat b_i\left|\psi_i(t)\right|,
\end{align*}
where $c_{1i} = (1/\underline \tau_i+2\overline c_i\overline v_i/\underline m_i)(k_{3i}+h_{2i}/h_{1i}+1/\kappa_{2i})h_{2i}\delta
         +2\overline c_i\overline v_i\overline a_i/(\underline m_i\underline \tau_i)+2\overline c_i\overline a_i^2/\underline m_i+ \sigma_{2i}
         +\overline c_{bi}[(h_{2i}^2+(k_{2i}+k_{3i})h_{2i}^2/h_{1i}+|k_{3i}^2-h_{2i}^2/h_{1i}^2|+|h_{2i}^2/h_{1i}^2-1/\kappa_{2i}^2|)\delta+\alpha_{4i}/\kappa_{2i}]$,
$c_{2i}= 1/\underline \tau_i+2\overline c_i\overline v_i/\underline m_i+\overline c_{bi}k_{3i}$.
This together with $\dot q_i(t)=w_i(t)$ and Lemma 1 leads to $|e_{1i}(t)|\leq\overline e_{1i},\ \forall\ t\geq0$.
From (\ref{dV_i}), noting that $|e_{1i}(t)|\leq\overline e_{1i}$, $|\dot \alpha_{1i}(t)|\leq \alpha_{3i}$ and $|\dot \alpha_{2i}(t)|\leq\alpha_{4i},\ \forall\ t\geq0$, we obtain
\begin{align}\label{dV_i1}
&\dot V_i(t)\nonumber\\
\leq&-\left(k_{1i}-1/2\right)e^2_i(t)-\left(k_{2i}-1/2\right)z^2_{1i}(t)\nonumber\\
           &-\left[k_{3i}-\overline e_{1i}^2/(2h_{2i}^2\xi_i)\right]z_{2i}^2(t)\nonumber\\
           &-\left[1/\kappa_{1i}-h_{1i}^2/2- \alpha^2_{3i}/(2\xi_i)\right]\eta^2_{1i}(t)\nonumber\\
&-\left[1/\kappa_{2i}-h_{2i}^2/(2h_{1i}^2)- \alpha_{4i}^2/(2\xi_i)\right]\eta_{2i}^2(t)+3\xi_i/2.
\end{align}
Denote $\rho_i = \min\{k_{1i}-1/2,\ k_{2i}-1/2, \ k_{3i}-\overline e_{1i}^2/(2h_{2i}^2\xi_i), \\ \ 1/\kappa_{1i}-h_{1i}^2/2- \alpha^2_{3i}/(2\xi_i),\ 1/\kappa_{2i}-h_{2i}^2/(2h_{1i}^2)- \alpha_{4i}^2/(2\xi_i)\}$. This together with (\ref{dV_i1}) leads to
\begin{align}\label{dV_i2}
\dot V_i(t)&\leq-2\rho_iV_i(t)+3\xi_i/2.
\end{align}
From (\ref{dV_i2}), we get
\begin{align}\label{dV_i3}
 V_i(t)&\leq V_i(0)e^{-2\rho_it}+3\xi_i(1-e^{-2\rho_it})/(4\rho_i).
\end{align}
By (\ref{beta_1i}) and (\ref{beta_2i}), we know $\eta_{1i}(0)=0$ and $\eta_{2i}(0)=0$, which leads to $V_i(0)=e_i^2(0)/2+z_{1i}^2(0)/2+z_{2i}^2(0)/2$.
By (\ref{z_1i})$-$(\ref{beta_2i}), (\ref{alpha_1i}) and (\ref{alpha_2i}), we know $z_{1i}(0)=(-v_{d,i}(0)-k_{1i}e_i(0))/h_{1i}$ and $z_{2i}(0) = (a_i(0)-k_{2i}v_{d,i}(0)-(k_{1i}k_{2i}+h_{1i}^2)e_i(0))/h_{2i}$.
This together with $V_i(0)=e_i^2(0)/2+z_{1i}^2(0)/2+z_{2i}^2(0)/2$ leads to
\begin{align}\label{V_i0}
V_i(0)=(A_{0i}e_i^2(0)+B_{0i}e_i(0)+C_{0i})/2.
\end{align}
From (\ref{V_i0}), we know that if
\begin{align*}
|e_i(0)|\leq\min\{|(-B_{0i}+(B_{0i}^2-4A_{0i}C_{0i})^{1/2})/(2A_{0i})|,\nonumber\\
|(- B_{0i}-(B_{0i}^2-4A_{0i}C_{0i})^{1/2})/(2A_{0i})|\}
\end{align*}
then $V_i(0)\leq \delta^2/2$,
where $A_{0i}=1+k_{1i}^2/h_{1i}^2+(k_{1i}k_{2i}+h_{1i}^2)^2/h_{2i}^2$, $B_{0i}=2k_{1i}v_{di}(0)/h_{1i}^2-2(a_i(0)-k_{2i}v_{di}(0))(k_{1i}k_{2i}+h_{1i}^2)/h_{2i}^2$, $C_{0i}=v_{di}^2(0)/h_{1i}^2+(a_i(0)-k_{2i}v_{di}(0))^2/h_{2i}^2-\delta^2$.

From (\ref{parameters1}) and (\ref{parameters2}),  we know $\rho_i\geq3\xi_i/(2\epsilon^2)$. This together with $\epsilon\leq\delta$ leads to $\rho_i\geq3\xi_i/(2\delta^2)$.
From (\ref{dV_i3}), noting that $V_i(0)\leq \delta^2/2$ and $\rho_i\geq3\xi_i/(2\delta^2)$, we know $V_i(t)\leq\delta^2/2,\ \forall\ t>0$, which means $\max_{i=1,2,...,N}\sup_{t>0}|e_i(t)|\leq \delta$, so the string stability is guaranteed.

From (\ref{dV_i3}) and $\rho_i\geq3\xi_i/(2\epsilon^2)$, we know $\limsup_{t\to \infty}V_i(t)\leq (3\xi_i)/(4\rho_i)\leq\epsilon^2/2$,
which means $\limsup_{t\to \infty}|e_i(t)|\leq \epsilon$, so the closed-loop stability is guaranteed.

From (\ref{u_i2}), noting that $|e_{1i}(t)|\leq\overline e_{1i}$,  we know
\begin{align}\label{absdu_i}
|\dot u_i(t)|\leq l_i|\psi_i(t)|+B_i,
\end{align}
where $B_i = h_{2i}[(l_i+k_{3i})\overline e_{1i}/h_{2i}+(h_{2i}+(k_{2i}+k_{3i})h_{2i}/h_{1i}+|k_{3i}^2-h_{2i}^2/h_{1i}^2|+|h_{2i}^2 /h_{1i}^2-1/\kappa_{2i}^2|+\alpha_{4i}/\kappa_{2i})\delta ]/\hat b_i$.
From (\ref{ET1}), we know $\psi_i(t)=\gamma_i(t)-u_i(t)=u_i(t_k^i)-u_i(t)=-\int_{t_k^i}^{t} \dot u_i(\tau)\, d\tau,\ t\in[t_k^i,t_{k+1}^i)$.
This together with (\ref{absdu_i}) leads to $|\psi_i(t)|\leq\int_{t_k^i}^{t} |\dot u_i(\tau)|\, d\tau \leq\int_{t_k^i}^{t} l_i\left|\psi_i(\tau)\right|+B_i \, d\tau <\int_{t_k^i}^{t} l_iM_i+B_i \, d\tau=(t-t_k^i)(l_iM_i+B_i),\ t\in[t_k^i,t_{k+1}^i)$. Then we know that $|\psi_i(t)|<M_i$ when $t=t_k^i+M_i/(l_iM_i+B_i)$. According to (\ref{ET1}) and (\ref{ET2}), we get $t_{k+1}^i-t_k^i>M_i/(l_iM_i+B_i)$, so there exists a positive low bound of the time
interval of the event triggering, and the Zeno behavior is avoided under the designed event-trigged mechanism.\qed
\end{appendices}
\vspace{-2mm}
\bibliographystyle{model5-names}        % Include this if you use bibtex
\bibliography{LLG2022Automatica-arXiv-V1-2022.09.16}           % and a bib file to produce the
% bibliography (preferred). The
% correct style is generated by
% Elsevier at the time of printing.

%\begin{thebibliography}{99}     % Otherwise use the
% thebibliography environment.
% Insert the full references here.
% See a recent issue of Automatica
% for the style.
  %\bibitem[Heritage, 1992]{Heritage:92}
%     (1992) {\it The American Heritage.
%     Dictionary of the American Language.}
%     Houghton Mifflin Company.
%  \bibitem[Able, 1956]{Abl:56}
%     B.~C.~Able (1956). Nucleic acid content of macroscope.
%     {\it Nature 2}, 7--9.
%  \bibitem[Able {\em et al.}, 1954]{AbTaRu:54}
%     B.~C. Able, R.~A. Tagg, and M.~Rush (1954).
%     Enzyme-catalyzed cellular transanimations.
%     In A.~F.~Round, editor,
%     {\it Advances in Enzymology Vol. 2} (125--247).
%     New York, Academic Press.
%  \bibitem[R.~Keohane, 1958]{Keo:58}
%     R.~Keohane (1958).
%     {\it Power and Interdependence:
%     World Politics in Transition.}
%     Boston, Little, Brown \& Co.
%  \bibitem[Powers, 1985]{Pow:85}
%     T.~Powers (1985).
%     Is there a way out?
%     {\it Harpers, June 1985}, 35--47.

%\end{thebibliography}

\end{document}